\documentclass[aps,onecolumn,superscriptaddress,preprintnumbers,amsmath,amssymb,preprint]{revtex4}

\usepackage{graphicx,color,dcolumn,booktabs,bm}
\usepackage{makecell}
\usepackage{indentfirst}
\usepackage{cases}
\usepackage[colorlinks,linkcolor=blue,anchorcolor=green,citecolor=blue]{hyperref}
\usepackage{tikz} 
\usepackage{tikz-feynman}
\tikzfeynmanset{compat=1.1.0} 

\usetikzlibrary{decorations.pathmorphing, arrows.meta} 

\definecolor{nicered}{rgb}{.7,.1,.1}
\definecolor{nicegreen}{rgb}{.1,.5,.1}
\definecolor{darkblue}{rgb}{0,0,.5}
\hypersetup{colorlinks, citecolor=nicegreen,linkcolor=nicered, urlcolor=darkblue}

\begin{document}


\title{Triple top baryon $\Omega_{ttt}$}

\author{Zi-Qi Zhu}
\affiliation{Department of Physics, Fudan University, Shanghai 200433, China}

\author{Chang Xiong}
\affiliation{School of Physics, Beihang University, Beijing 100191, China}

\author{Yu-Jie Zhang}
\thanks{zyj@buaa.edu.cn}
\affiliation{School of Physics, Beihang University, Beijing 100191, China}

\today

\begin{abstract}
The recent observation of toponium by CMS and ATLAS has renewed interest in top quark bound states. In this work, we present an exploratory but quantitative study of the hypothetical triple-top baryon, denoted as $\Omega_{ttt}$, the only baryon that is governed by ultraviolet freedom. Using a variational method with an effective potential of $ttt$ that includes QCD, Higgs, and QED contributions, we estimate its mass to be around 514 GeV with a binding energy of about 4 GeV. We further discuss its possible production at future high-energy colliders, finding that the cross sections are extremely suppressed. The dominant weak decay channel is identified as $\Omega_{ttt}\to W^+W^+W^+bbb$, leading to complex multi-lepton and multi-jet final states. Our analysis, though approximate, demonstrates the distinctive features of  $\Omega_{ttt}$ compared with other triply-heavy baryons such as $\Omega_{ccc}$ and  $\Omega_{bbb}$, and may serve as a starting point for more refined approaches, including lattice QCD or effective field theory. This work highlights both the theoretical challenges and the potential opportunities in probing the strong interaction at unprecedented mass scales.\end{abstract}

\maketitle
\nopagebreak

\section{Introduction}
\label{sec:intro}

The recent observation of toponium, a bound state of $t\bar{t}$, by CMS~\cite{CMS:2025kzt} and ATLAS~\cite{ATLAS:2025mvr} with an observed significance above $5\sigma$ has renewed interest in exotic heavy-quark bound states. In particular, it naturally motivates the study of the hypothetical triple-top baryon $\Omega_{ttt}$. Triply heavy baryons have been investigated for several decades~\cite{Bjorken:1985ei}, and although no experimental evidence has yet been reported, a variety of theoretical approaches have explored their spectroscopy, production, and decay properties. These methods include potential models~\cite{Bali:2000gf,Brambilla:2005yk,Brambilla:2009cd,Brambilla:2013vx,Koma:2017hcm,Assi:2025ysr}, quark and diquark models~\cite{Meng:2023jqk,Xie:2024lfo,Thakkar:2016sog,Anwar:2017toa,Liu:2019vtx,Yang:2019lsg}, QCD sum rules~\cite{Wang:2011ae,Wang:2020avt}, variational and Hartree--Fock methods~\cite{Jia:2006gw,Llanes-Estrada:2011gwu,deArenaza:2024dhe,Alonso-Valero:2024jim}, Faddeev equations~\cite{Sanchis-Alepuz:2011xjl,Radin:2014yna}, Gaussian expansions~\cite{
Zhu:2024hgm,
Ma:2025rvj,Ma:2025rvj,Wu:2025wvv,Yang:2025wqo,Yang:2025oih}, diffusion Monte Carlo~\cite{Ma:2022vqf,Ma:2023int}, and lattice QCD calculations~\cite{Brown:2014ena,Li:2022vbc,Zhou:2025fpp}. These complementary tools provide a framework for exploring the systematics of multi-heavy baryons.

The top quark, with mass $m_t = 172.57\pm 0.29~\mathrm{GeV}$~\cite{ParticleDataGroup:2024cfk} and width $\Gamma_t \simeq 1.3~\mathrm{GeV}$~\cite{Chen:2023osm,Chen:2023dsi}, is the heaviest known elementary particle. Its very short lifetime, $\tau_t \simeq 5\times 10^{-25}$\,s~\cite{Bigi:1986jk}, is much shorter than the typical hadronization time scale $1/\Lambda_{\mathrm{QCD}}\sim 10^{-23}$\,s. As a result, conventional toponium states were long thought not to form before decay. 
However, more refined analyses~\cite{Fu:2025yft,Fu:2025zxb,Xiong:2025iwg} suggest that the Bohr radius and formation timescales of $t\bar{t}$ and multi-top systems may allow bound states to form within the top lifetime. For the triple-top baryon, since the potential between top quarks is weaker than that in toponium~\cite{Jia:2006gw}, the corresponding hadronization timescale is somewhat longer. Nevertheless, it remains shorter than the top-quark lifetime, implying that $\Omega_{ttt}$ could still form before decay. Furthermore, its dominant weak decay channel into $W^+W^+W^+bbb$ is essentially determined by the constituent top decays and is therefore largely insensitive to the details of hadronization.
Moreover, the characteristic scale $m_t v^2 \sim 3$~GeV~\cite{Fu:2025yft,Fu:2025zxb,Xiong:2025iwg} is well above $\Lambda_{\mathrm{QCD}}$, implying that $\Omega_{ttt}$ is a weakly coupled bound state amenable to perturbative treatments~\cite{Brambilla:2005yk,Jia:2006gw}.

The phenomenology of top-quark bound systems has been extensively discussed in the literature, including toponium production and decay~\cite{Hirata:1979xx,Fadin:1987wz,Djouadi:2024lyv,Francener:2025tor,Bai:2025buy}, and possible ``topped'' mesons and baryons~\cite{Zhang:2025xxd,Zhang:2025fdp,Luo:2025psq}. On the experimental side, four-top production has already been observed at the LHC~\cite{ATLAS:2023ajo,CMS:2023ftu} and studied theoretically at NLO and beyond~\cite{Frederix:2017wme,vanBeekveld:2022hty,Cao:2019qrb}. Even more extreme processes, such as six-top production, have been proposed~\cite{Deandrea:2014raa,Han:2018hcu,Bernreuther:2025uwp}. These results demonstrate the increasing feasibility of probing top-rich final states at future very high energy colliders.

Three-body bound states are notoriously challenging to treat exactly, particularly in the heavy-quark sector. In this work we adopt a variational approach with an effective potential to provide an exploratory yet quantitative estimate of the mass, production mechanisms, and decay properties of the triple-top baryon $\Omega_{ttt}$. Our results establish a theoretical baseline for future studies and help to clarify how the strong interaction behaves at the highest mass scales accessible within the Standard Model.
 
\section{Mass and Properties of $\Omega_{ttt}$}
\label{sec:mass}

Since the triple-top baryon is composed of three identical spin-$\frac{1}{2}$ fermions, the Pauli exclusion principle requires their color wave function to be antisymmetric. Consequently, the spin wave function must be symmetric, leading to a total spin of $J = \frac{3}{2}$. The ground state of the triple-top system, denoted as $\Omega_{ttt}$, is therefore expected to have quantum numbers $J^P = \frac{3}{2}^+$, analogous to the $\Omega$ baryon. 

\begin{figure}[t]
  \centering
  \begin{tikzpicture}
    \draw[-, thick] (0.,0.) -- (4,3);  
    \draw[-, thick] (0.0,0.0) -- (-2,4);  
    \draw[-, thick] (-2,4) -- (4,3);  
    \node at (2.2,1.0) {$r_1$};
    \node at (-1.5,2.0) {$r_2$};
    \node at (1.0,3.0) {$r_{12}$};
    \shade (0,0) circle [radius=0.40cm]; \node at (0,0) {$t$};
    \shade (4,3) circle [radius=0.40cm]; \node at (4,3) {$t$};
    \shade (-2,4) circle [radius=0.40cm]; \node at (-2,4) {$t$};
    \node at (5.0,0.0) {$\langle r_{tt}\rangle=0.016~\mathrm{fm}$};
  \end{tikzpicture}
  \caption{\label{Fig:Omegattt}Schematic diagram of the $\Omega_{ttt}$ structure.}
\end{figure}
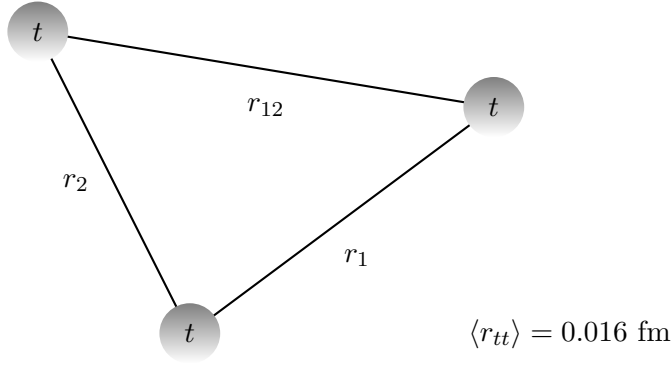

Because $\Omega_{ttt}$ is a weakly coupled state, its dynamics can be approximately described by the non-relativistic Schr\"odinger equation~\cite{Brambilla:2005yk,Jia:2006gw}, with a typical quark velocity $v \sim 0.1$~\cite{Fu:2025yft,Fu:2025zxb,Xiong:2025iwg}. The Hamiltonian reads
\begin{align}
H = -\frac{1}{2}\sum_{i=1}^3 \frac{\nabla_i^2}{m_t} + V_{ttt},
\end{align}
where $V_{ttt}$ is the effective potential of $ttt$. After separating the center-of-mass (c.m.) motion, the quark coordinates can be written as
\begin{align}
\vec{x}_1 &= \vec{X} + \frac{2}{3}\vec{r}_1 - \frac{1}{3}\vec{r}_2, \nonumber\\
\vec{x}_2 &= \vec{X} - \frac{1}{3}\vec{r}_1 + \frac{2}{3}\vec{r}_2, \nonumber\\
\vec{x}_3 &= \vec{X} - \frac{1}{3}\vec{r}_1 - \frac{1}{3}\vec{r}_2,
\end{align}
with
\begin{align}
\vec{X} &= \frac{\vec{x}_1+\vec{x}_2+\vec{x}_3}{3},  \nonumber\\
\vec{r}_1 &= \vec{x}_1 - \vec{x}_3,  \nonumber\\
\vec{r}_2 &= \vec{x}_2 - \vec{x}_3,
\end{align}
where $\vec{X}$ is the c.m. position and $\vec{r}_{1,2}$ denote the relative coordinates of $t_1$ and $t_2$ with respect to $t_3$ (see Fig.~\ref{Fig:Omegattt}). 

In the c.m. frame, the kinetic term becomes
\begin{align}
\frac{1}{2}\sum_{i=1}^3 \frac{\nabla_i^2}{m_t}
= -\frac{\nabla^2_{\vec{X}}}{6 m_t}
 - \frac{\nabla^2_{r_1}}{m_t}
 - \frac{\nabla^2_{r_2}}{m_t}
 - \frac{\nabla_{r_1}\cdot\nabla_{r_2}}{m_t},
\end{align}
and after removing the c.m. part, the Hamiltonian reduces to
\begin{align}
H = -\frac{\nabla^2_{r_1}}{m_t}
    -\frac{\nabla^2_{r_2}}{m_t}
    -\frac{\nabla_{r_1}\cdot\nabla_{r_2}}{m_t}
    + V_{ttt}.
\end{align}

The potential $V_{ttt}$ among the three top quarks is approximated as the sum of pairwise interactions~\cite{Jia:2006gw}:
\begin{align}
V_{ttt} = V_{tt}(r_1)+V_{tt}(r_2)+V_{tt}(r_{12}),
\end{align}
with
\begin{align}
V_{tt}(r) &= V^{\rm QCD}(r) + V^{\rm Higgs}(r) + V^{\rm QED}(r), \nonumber\\
V^{\rm QCD}(r) &= -\frac{\lambda_3}{r} + \sigma r, \qquad 
 \lambda_3 = 0.154 \pm 0.005,~ \sigma = 0.206~\text{GeV}^2, \nonumber\\
V^{\rm QED}(r) &= \frac{4\alpha}{9r}, \nonumber\\
V^{\rm Higgs}(r) &= -\frac{m_t^2}{4\pi v^2}\frac{e^{-m_H r}}{r}, \qquad v=246.22~\text{GeV}.
\end{align}
Here the color-triplet Coulomb coefficient $\lambda_3$ is half of the color-singlet Coulomb coefficient due to group factors~\cite{Fu:2025yft,Fu:2025zxb}.

To estimate the mass we adopt a variational ansatz~\cite{Jia:2006gw} with trial wave function
\begin{align}
\psi(r_1,r_2) = 
\frac{f(r_1)f(r_2)+f(r_{12})f(r_2)+f(r_1)f(r_{12})}{\sqrt{595}/9},
\end{align}
where $f(r)$ is a normalized hydrogen-like wave function with variational parameter $\Lambda$,
\begin{align}
f(r) = \frac{\Lambda^{3/2}}{\sqrt{\pi}}e^{-\Lambda r}.
\end{align}

The resulting energy expectation value is
\begin{align}
E(\Lambda) = \frac{1062}{595 m_t}\Lambda^2
 - \frac{2595}{952}\left(\lambda_3 - \frac{4}{9}\alpha\right)\Lambda
 + \frac{28127\sigma}{5712}\frac{1}{\Lambda}
 + E^{\rm Higgs}(\Lambda).
\end{align}
The $\Lambda^2$ and $\Lambda$ terms are given in Ref.~\cite{Jia:2006gw}, which correspond to  the kinetic energy and  Coulomb potential respectively.  The $\frac{1}{\Lambda}$ term arises from the linear potential $ \sigma r$.  
where $E^{\rm Higgs}(\Lambda)$ is small and not shown explicitly. 

Applying the variational principle, 
\begin{align}
\frac{{\rm d}}{{\rm d} \Lambda}E(\Lambda)=0.
\end{align}
Minimizing $E$ with respect to $\Lambda$ gives the numerical result
\begin{align}
\Lambda = 20.6~\text{GeV},
\end{align}
which is close to the inverse Bohr radius of toponium ($1/26.7$~GeV). 
The average of total kinetic energy  in the center-of-mass frame is 
\begin{align}
\langle T\rangle=4.40~{\rm GeV},
\end{align}
and the typical energy
\begin{align}
\langle m_t v^2\rangle=\frac{2}{3}\times4.40~{\rm GeV}=2.93~{\rm GeV}.
\end{align}
The corresponding expectation value of $tt$ distance
\begin{align}
\langle r_{tt} \rangle&=\frac{1}{12.6~\rm GeV}.
\end{align}
The wave function at the origin is found to be
\begin{align}
\psi(0,~0)&=3093~{\rm GeV}^3.
\end{align}
The binding energy is found to be
\begin{align}
E = -4.13 \pm 0.27~\text{GeV},
\end{align}
where the Higgs ($-0.115$~GeV) and linear potential ($+0.049$~GeV) contributions are included, and the uncertainties mainly reflect variations in $\lambda_3$.  The mass of $\Omega_{ttt}$ is estimated as
\begin{align}
m(\Omega_{ttt}) = 513.58 \pm 0.87_{m_t} \pm 0.23_{\lambda_3}~\text{GeV},
\end{align}
where the uncertainties mainly reflect variations in top quark on-shell mass.
This value is consistent with expectations for a weakly bound system and provides a quantitative baseline for comparison with future lattice QCD or effective field theory studies.

\section{Production of $\Omega_{ttt}$ at Colliders}

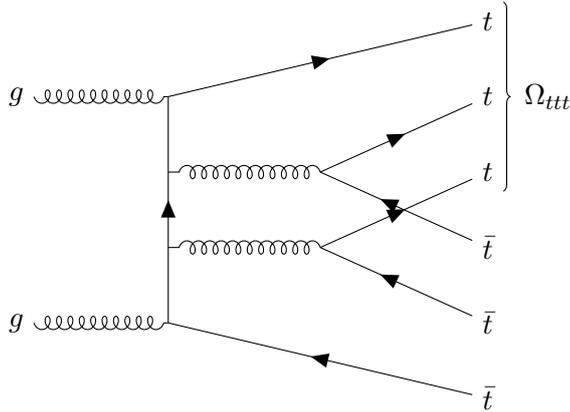
\begin{figure}[t]
\centering
\begin{tikzpicture}
\begin{feynman}
  \vertex (g1) {\(g\)};
  \vertex[below=3cm of g1] (g2) {\(g\)};

  \vertex[right=2cm of g1] (v1);

  \vertex[below=1cm of v1] (v2);
  \vertex[below=1cm of v2] (v3);
  \vertex[right=2cm of v2] (v5);
  \vertex[right=2cm of v3] (v6);

  \vertex[below=1cm of v3] (v4);
  \vertex[right=4cm of v1] (t2) {\(t\)};
  \vertex[above=1cm of t2] (t1) {\(t\)};
  \vertex[below=1cm of t2] (t3) {\(t\)};

  \vertex[below=1cm of t3] (at1) {\(\bar{t}\)};
  \vertex[below=1cm of at1] (at2) {\(\bar{t}\)};
  \vertex[below=1cm of at2] (at3) {\(\bar{t}\)};


  \diagram*{
    (g1) -- [gluon] (v1);
    (g2) -- [gluon] (v4);
    (v5) -- [gluon] (v2);
    (v6) -- [gluon] (v3);
    (at3) -- [fermion] (v4) -- [fermion] (v1)-- [fermion] (t1),
    (v5) -- [fermion] (t2),
    (v6) -- [fermion] (t3),
    (v5) -- [anti fermion] (at1),
    (v6) -- [anti fermion] (at2),
  };

  \draw [decoration={brace}, decorate] (t1.north east) -- (t3.south east)
    node [pos=0.5, right=4pt] {\(\Omega_{ttt}\)};
\end{feynman}
\end{tikzpicture}
\caption{Schematic Feynman diagram of $\Omega_{ttt}$ production via gluon fusion at hadron collider:
$gg \to t\bar t\,t\bar t\,t\bar t \to \Omega_{ttt} + \bar t\bar t\bar t$.}
\label{fig:feynmanOmega}
\end{figure}

The production of $\Omega_{ttt}$ at hadron colliders requires the creation of six top quarks, e.g.
\begin{align}
gg \to \Omega_{ttt} + \bar{t}+\bar{t}+\bar{t}.
\end{align}
Schematic Feynman diagram of $\Omega_{ttt}$ production via gluon fusion is shown in Fig.~\ref{fig:feynmanOmega}. The Electroweak and Higgs contributions are included too.

%
%
The production cross section can be estimated following Ref.~\cite{Chen:2011mb} as
\begin{align}\label{eq:omegatttX}
\sigma(pp\to \Omega_{ttt}/\bar{\Omega}_{ttt})
&\sim 2 \times 3! \times \frac{|\psi(0,0)|^2}{m_t^6}\,
\sigma(pp\to 3\, t\bar{t}) \nonumber\\
&\sim 4.3\times 10^{-6}\, \sigma(pp\to 3\, t\bar{t}),
\end{align}
where the factor of 2 accounts for the charge-conjugate state $\bar{\Omega}_{ttt}$ and the factor of $3!$ reflects the identical-quark symmetry. The dimensionless factor $\frac{|\psi(0,0)|^2}{m_t^6}$ corresponds to the three- body bound state effect.

As a validation, applying Eq.~(\ref{eq:omegatttX}) to $\Omega_{ccc}$ production at the LHC with $\sqrt{s}=14$~TeV and $|\eta|<2.5$ yields a cross section of $\sim 0.058$~nb, which is within a factor of two of the full calculation ($0.113$~nb) reported in Ref.~\cite{Chen:2011mb}. This agreement indicates that the approximation provides a reasonable order-of-magnitude estimate.

\begin{figure}[t]
  \centering
  \includegraphics[width=0.95\linewidth]{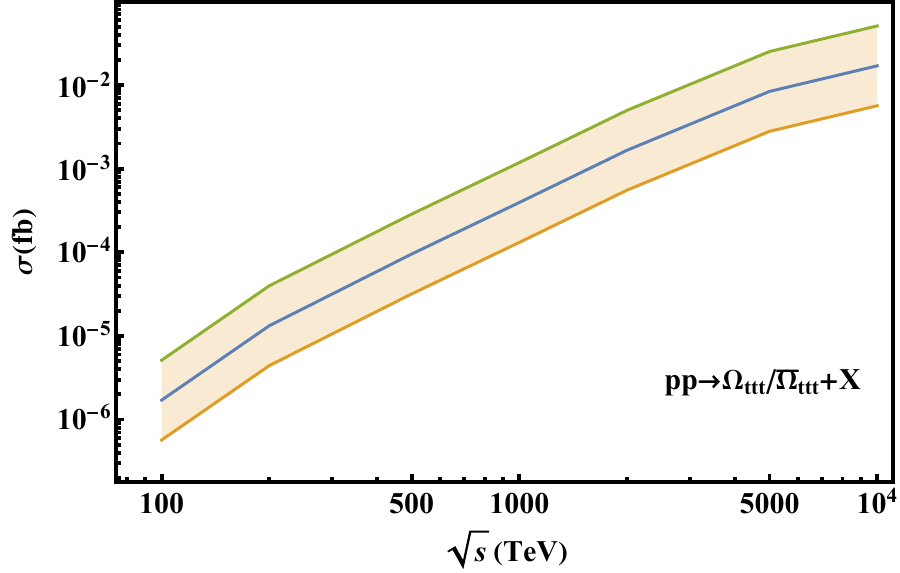}
  \caption{\label{Fig:ppOmegattt}
Estimated cross section of $pp\to \Omega_{ttt}/\bar{\Omega}_{ttt}$ as a function of $\sqrt{s}$ in TeV. Electroweak and Higgs contributions are included here. 
The error band, which arises primarily from scale and parton distribution function (PDF) uncertainties, spans from the central value multiplied by a factor of $1/4$ to that multiplied by a factor of 4.
}
\end{figure}

Using HELAC-Onia~2.0~\cite{Shao:2015vga}, we estimate the leading-order cross sections including Electroweak and Higgs contributions at $pp$ colliders  as
\begin{align}
\sigma(pp\to \Omega_{ttt}/\bar{\Omega}_{ttt}\,|\,\sqrt{s}=10^2~{\rm TeV}) &\sim 1.7\times 10^{-7}~\text{fb}, \nonumber\\
\sigma(pp\to \Omega_{ttt}/\bar{\Omega}_{ttt}\,|\,\sqrt{s}=10^3~{\rm TeV}) &\sim 3.9\times 10^{-4}~\text{fb}, \nonumber\\
\sigma(pp\to \Omega_{ttt}/\bar{\Omega}_{ttt}\,|\,\sqrt{s}=10^4~{\rm TeV}) &\sim 1.7\times 10^{-2}~\text{fb}.
\end{align}
The dominant production mechanism is gluon fusion, requiring the simultaneous creation of three $t\bar{t}$ pairs. The cross sections are orders of magnitude smaller than for toponium, implying that even at future colliders such as the FCC-hh~\cite{FCC:2025uan} or SPPC~\cite{Tang:2022fzs}, the observation of $\Omega_{ttt}$ would demand integrated luminosities far beyond foreseeable capabilities. The cross section of $pp\to \Omega_{ttt}/\bar{\Omega}_{ttt}$ as a function of $\sqrt{s}$ in TeV is shown in Fig.~\ref{Fig:ppOmegattt}.

\begin{figure}[t]
  \centering
  \includegraphics[width=0.95\linewidth]{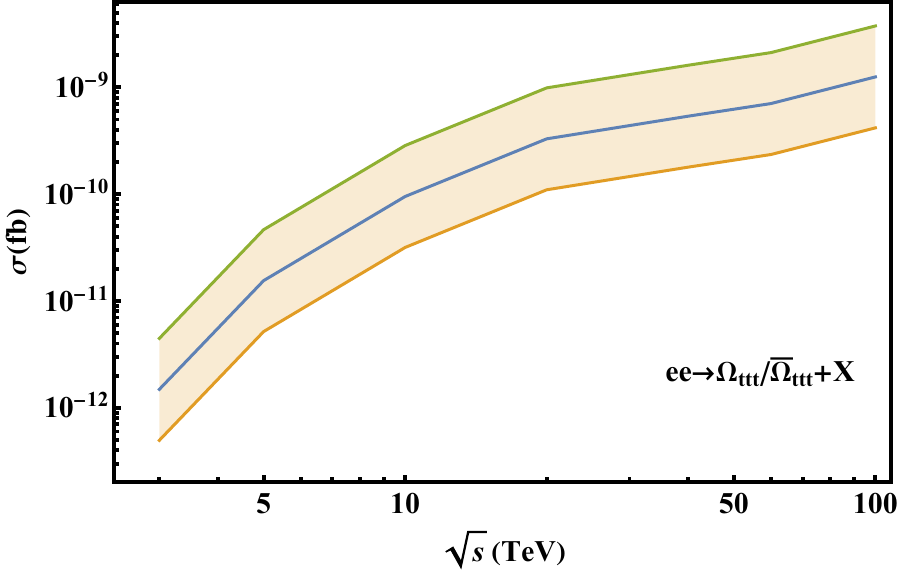}
  \caption{\label{Fig:eeOmegattt}
Estimated cross section of $e^+e^-\to \Omega_{ttt}/\bar{\Omega}_{ttt}$ as a function of $\sqrt{s}$ in TeV. Electroweak and Higgs contributions are included here. The error band, which arises primarily from scale uncertainties, spans from the central value multiplied by a factor of $1/3$ to that multiplied by a factor of 3.}
\end{figure}

We also estimate the leading-order cross sections at lepton $e^+e^-$~\cite{FCC:2025lpp,CEPCStudyGroup:2023quu,ILCInternationalDevelopmentTeam:2022izu,CLICdp:2018cto} or $\mu^+\mu^-$~\cite{Accettura:2023ked,InternationalMuonCollider:2024jyv} colliders  as
\begin{align}
\sigma(e^+e^-\to \Omega_{ttt}/\bar{\Omega}_{ttt}\,|\,\sqrt{s}=10~{\rm TeV}) &\sim 9.5\times 10^{-11}~\text{fb}, \nonumber\\
\sigma(e^+e^-\to \Omega_{ttt}/\bar{\Omega}_{ttt}\,|\,\sqrt{s}=40~{\rm TeV}) &\sim 5.4\times 10^{-10}~\text{fb}, \nonumber\\
\sigma(e^+e^-\to \Omega_{ttt}/\bar{\Omega}_{ttt}\,|\,\sqrt{s}=100~{\rm TeV}) &\sim 1.3\times 10^{-9}~\text{fb}.
\end{align}
The cross sections are orders of magnitude smaller than from hadron colliders. The cross section of $e^+e^-\to \Omega_{ttt}/\bar{\Omega}_{ttt}$ as a function of $\sqrt{s}$ in TeV is shown in Fig.~\ref{Fig:eeOmegattt}.

\section{Decay}

Neglecting small relativistic corrections, the total width of $\Omega_{ttt}$ is estimated as~\cite{Fu:2025yft,Fu:2025zxb}
\begin{align}
\Gamma_{\rm total}(\Omega_{ttt}) \simeq 3\Gamma_t \approx 3.93~\text{GeV}.
\end{align}
Since $\Omega_{ttt}$ consists of three top quarks, its primary decay proceeds through the standard weak decay $t\to W^+b$, yielding
\begin{align}
\Omega_{ttt} \to (W^+b)+(W^+b)+(W^+b).
\end{align}
This decay is essentially insensitive to hadronization details, as the weak decay of the top quark dominates. A rare transition into a triply-bottom baryon is also possible,
\begin{align}
\Omega_{ttt} \to W^+W^+W^+ + \Omega_{bbb},
\end{align}
with branching ratio
\begin{align}
{\rm Br}(\Omega_{ttt}\to W^+W^+W^+ + \Omega_{bbb}) \sim \frac{m_b^6}{m_t^6}\sim 10^{-9}.
\end{align}

Subsequent $W$ decays into leptons or hadrons produce complex final states with multiple jets and leptons. A characteristic experimental signature would be three $b$-jets and three reconstructed $W$ bosons, with an invariant mass peak near $\sim 514$~GeV. However, this channel faces overwhelming backgrounds from multi-top production.

The main properties of $\Omega_{ttt}$ are summarized in Table~\ref{tab:omegatttToponium}, together with a comparison to spin-triplet ($J_t$) and spin-singlet ($\eta_t$) toponium.

\begin{table}[htbp]
\caption{Comparison of the main properties of the triple-top baryon $\Omega_{ttt}$ and toponium states ($J_t$, $\eta_t$)~\cite{Fu:2025yft,Fu:2025zxb}.} 
\label{tab:omegatttToponium} 
\begin{center}
\begin{tabular}{cccc}
\hline
Particle&   $\Omega_{ttt}$  & $J_t$ & $\eta_t$\\ \hline
$J^P$ or $J^{PC}$ & $\frac{3}{2} {}^+$ &$1^{--}$&$0^{-+}$\\
Mass (GeV)& 513.58& 341.02&341.02 \\
Binding Energy (GeV)&4.13&4.12&4.12\\
Typical Scale $m_t v^2$  (GeV)&2.93&4.12&4.12\\
$\langle r_{tt} \rangle$ ~ (fm)&0.016&0.011&0.011\\
$1/\langle r_{tt} \rangle$ ~ (GeV)&12.6&17.8&17.8\\
Production Process &$pp\to \Omega_{ttt}+3\bar{t}$ &$e^+e^-\to J_t$&$pp\to \eta_t+X$\\
Total Width   (GeV)&3.93&2.604&2.615\\ 
Typical non-$Wb$ decays &~ $\Omega_{bbb}+3W$ ($\sim 10^{-9}$) ~& ~$WW,~b\bar b$ ($\sim  10^{-3}$)~ &~ $gg,~ZH$ ($\sim 10^{-3}$) \\
 \hline 
\end{tabular}
\end{center}
\end{table}

\section{Summary and Outlook}

The production of the triple-top baryon $\Omega_{ttt}$ is expected to be extremely suppressed, rendering its direct observation highly challenging at present or near-future colliders. The main difficulties arise from the tiny cross section and the large Standard Model backgrounds in multi-top final states.

Nevertheless, our analysis establishes a theoretical baseline for $\Omega_{ttt}$ properties, including its mass, binding energy, production cross sections, and dominant decay channels. The results highlight distinctive features of $\Omega_{ttt}$ compared with other triply heavy baryons. Importantly, the weak-decay nature of the top quark ensures that the $\Omega_{ttt}\to 3W+3b$ final state is largely unaffected by hadronization, providing a clean theoretical prediction.

Future progress will likely come from improved theoretical tools, such as lattice QCD or effective field theories, to refine mass and cross-section predictions. Although discovery prospects remain remote, the study of $\Omega_{ttt}$ offers a unique window into QCD at the highest mass scales and serves as a benchmark for future explorations of multi-heavy-quark dynamics.


\acknowledgments

We thank Dr. Jing-Hang Fu and Hui-min Yang for valuable discussions.


\providecommand{\href}[2]{#2}\begingroup\raggedright\endgroup

\end{document}